\documentclass[review]{elsarticle}

\usepackage{lineno,hyperref}
\modulolinenumbers[5]

\usepackage[version=3]{mhchem} 
\usepackage{amsfonts}
\usepackage{multirow}
\usepackage{subfig}
\usepackage{bm}

\usepackage{hyperref}

\usepackage{comment}
\usepackage{color}
\specialcomment{notes}{\begingroup\sffamily\color{red}}{\endgroup}

\usepackage{listings}
\usepackage{sistyle}
\SIthousandsep{,}

\begin{document}

\begin{frontmatter}

\title{Technical Report -- Comparison of Direct Finite Element Simulation with Actuator Line Models and Vortex Models for Simulation of Turbulent Flow Past a Vertical Axis wind Turbine}

\author[KTH]{Van-Dang Nguyen}
\address[KTH]{Department of Computational Science and Technology, KTH Royal Institute of Technology, Stockholm, Sweden}
\ead{vdnguyen@kth.se}

\author[KTH]{Johan Jansson} 
\ead{jjan@kth.se}
\author[UU]{Anders Goude}
\ead{anders.goude@angstrom.uu.se}
\author[KTH]{Johan Hoffman}
\ead{jhoffman@kth.se}

\address[UU]{{\AA}ngstr{\"o}m Laboratory, Uppsala University, Uppsala, Sweden}


\begin{abstract}
We compare three different methodologies for simulation of turbulent flow past a vertical axis wind turbine: (i) full resolution of the turbine blades in a Direct Finite Element Simulation (DFS), (ii) implicit representation of the turbine blades in a 3D Actuator Line Method (ALM), and (iii) implicit representation of the turbine blades as sources in a Vortex Model (VM). The integrated normal force on one blade is computed for a range of azimuthal angles, and is compared to experimental data for the different tip speed ratios, 2.55, 3.44 and 4.09. 
\end{abstract}

\begin{keyword}
vertical axis wind turbine, direct finite element simulation, actuator line method, vortex method, FEniCS-HPC
\end{keyword}

\end{frontmatter}

\section{Introduction}

Simulation models for vertical axis wind turbines (VAWTs) can be divided into three classes of methods. The first class is based on computational solution of the Navier-Stokes equations, also referred to as computational fluid dynamics (CFD), e.g. by a finite element method (FEM) or a finite volume method (FVM). The second class is based on computational solution of the vorticity equation, referred to as vortex models. The third class of methods is based on the momentum conservation principle, for example, the double multiple streamtube model.

\section{Methods}

In this report we focus of the CFD and vortex models. We compare one CFD method 
based on full resolution of the turbine blades, Direct Finite Element Simulation (DFS), one CFD method based on an implicit representation of the turbine blades, Actuator Line Model (ALM), and 2D/3D vortex models (VM).  

\subsection{Direct Finite Element Simulation}

In a Direct Finite Element Simulation (DFS) the airflow around the VAWT is modeled by the Navier-Stokes equations. For incompressible flow, the equations read
\begin{equation}
\begin{aligned}
\frac{\partial u}{\partial t} + ({u} \cdot \nabla) \,{u} - \nu \Delta {u} + \nabla p &= {f}, \quad \text{in $\Omega\times I$,} \\
\nabla \cdot {u} &= 0, \quad \text{in $\Omega\times I$,}\\
{u}(\cdot,0) &= {u}_0, \quad \text{in $\Omega$,}
\end{aligned}
\label{eq:primal-problem}
\end{equation}
where ${u}$ is the velocity, $p$ pressure and ${f}$ a given body force. $\Omega \subset \mathbb{R}^3$ is a spatial domain with 
boundary $\Gamma$, and $I=[0, T]$ a time interval. 

For a moving or deforming domain, we use an Arbitrary Lagrangian-Eulerian (ALE) method \cite{donea2004arbitrary}, which is 
based on the introduction of a separate set of reference coordinates. Often we let these reference coordinates trace the deformation of 
the finite element mesh, described by the mesh velocity $\beta$. In an ALE method, the convection term is modified to take the mesh velocity into account, 
which gives the modified Navier-Stokes equations on ALE form,
\begin{equation}
\begin{aligned}
\frac{\partial u}{\partial t} + \Bigl(({u}-{\beta}) \cdot \nabla\Bigl){u} - \nu \Delta {u} + \nabla p &= {f}, \quad \text{in $\Omega\times I$,} \\
\nabla \cdot {u} &= 0, \quad \text{in $\Omega\times I$,}\\
{u}(\cdot,0) &= {u}_0, \quad \text{in $\Omega$.}
\end{aligned}
\label{eq:primal-ale-problem}
\end{equation}

The DFS-ALE used to discretize the flow around the VAWT \cite{NGUYEN2019238} was developed in the framework of 
a Galerkin least-squares space-time finite element method (GLS) \cite{2006CMAME.195.1709H}, corresponding to a DFS method which can simulate both 
laminar and turbulent flow \cite{HOFFMAN201560}
Let $0=t_0<t_1<\dots<t_N=T$ be a time partition
associated with the time intervals $I_n=(t^{n-1}, t^n]$ of length
$k_n=t^n-t^{n-1}$. We denote the finite element space of continuous piecewise linear functions by $Q_h$, with the derived spaces 
$Q_{h,0}=\{q\in Q_h: q(x)=0,~x\in \Gamma\}$ and $V_h=[Q_{h,0}]^3$. The DFS-ALE method with least-squares stabilization is stated as: For all time intervals $I_{n}$, find $(U^{n}_h, P^{n}_h)\in V_h\times Q_h$ such that
\begin{multline*}
\Bigl(\frac{U^{n}_h-U^{n-1}_h}{k_{n}}+\bigl((\bar{U}^{n}_h-\beta_h)\cdot \nabla\bigl)\bar{U}^{n}_h,v_h\Bigl)\\+\Bigl( \nu \nabla \bar{U}^{n}_h, \nabla v_h\Bigl)-\Bigl( P^{n}_h, \nabla \cdot v_h\Bigl) +\Bigl( \nabla\cdot \bar{U}^n_h,q_h \Bigl) \\
+ SD_\delta\Bigl(\bar{U}^n_h, P^{n}_h; v_h, q_h \Bigl) = \bigl(f,v_h\bigl)
\end{multline*}
for all test functions $(v_h,q_h)\in V_h\times Q_h$, where $\bar{U}^n_h=\frac{U^{n}_h+U^{n-1}_h}{2}$, and $(U^{n}_h, P^{n}_h)$ is a numerical approximation of $(u, p)$ at $t=t_n$, and
with stabilization term
\begin{multline*}
SD_{\delta}^{n} (\bar{U}^n_h, P^n_h; v, q) :=
\Biggl( \delta_1\,\Bigl(  \bigl(\bar{U}^n_h-\beta_h\bigl) \cdot \nabla \bar{U}^n_h + \nabla P^n_h - f_n \Bigl) \,,\, \Bigl(\bar{U}^n_h-\beta_h\Bigl) \cdot \nabla v_h + \nabla q_h \Biggl) \\
+ \left( \delta_2\, \nabla \cdot \bar{U}^n_h\,,\, \nabla \cdot v_h \right). \hspace{5.8cm}
\label{eq:UCmodel_stab}
\end{multline*}
Here $\delta_1$ and $\delta_2$ are given stabilization parameters:
\begin{equation*}
\delta_1 = C_1 \Bigl(k_n^{-2} + |U^{n-1}_h-\beta_h|^2 h_n^{-2}\Bigl)^{-1/2},\quad\delta_2 =C_2\,|U^{n-1}_h|\,h_n.
\end{equation*}
We note that under a CFL condition, i.e $k_n= \frac{C_k\,h_n}{|U_h^{n-1}-\beta_h|}$, $\delta_1$ is simplified to $\delta_1=\bar{C}_1 \frac{h_n}{|U_h^{n-1}-\beta_h|}$. The Navier-Stokes equations are solved directly in weak forms without using any subgrid or turbulence models. Three parameters $\bar{C}_1, C_2,$ and $C_k$ need to be tuned during the simulations.

The method was implemented in the FEniCS-HPC platform \cite{Nguyen1105292, NGUYEN2019238, fenics-hpc:www}. The solver and the wind turbine data were published as an open-source package at 
\begin{center}
    \url{https://github.com/van-dang/VAWT-Cloud}.
\end{center}
The simulations can be performed on the Cloud using the Singularity container technology \cite{10.1371/journal.pone.0177459} as described in the following instruction
\begin{center}
    \url{https://github.com/van-dang/VAWT-Cloud/blob/master/README.md}
\end{center}
\subsection{Actuator Line Method}
The actuator line model (ALM) is based on the classical blade element model (BEM) theory coupled to a solver for the governing three-dimensional Navier-Stokes equations. The ALM \cite{669e7831503c4598a23340a8b0c5f56a} divides the blades in lines of elements which have a two-dimensional airfoil behavior, using tabulated lift and drag coefficients.

Using the Large Eddy Simulation (LES) approach for predicting turbulence effects based on an incompressible fluid, we consider the filtered Navier-Stokes equations, 
\begin{equation}
\begin{aligned}
\frac{\partial {\bar{u}}}{\partial t} + (\bar{u} \cdot \nabla)\, \bar{u} &= \nabla\cdot{\Bigl((\nu+\nu_{\rm SGS})\, \nabla \bar{u}\Bigl)} - \nabla \bar{p} - {f}, \quad \text{in $\Omega\times I$,} \\
\nabla \cdot {\bar{u}} &= 0, \quad \text{in $\Omega\times I$,}
\end{aligned}
\label{eq:filtered-NSE}
\end{equation}
where $\bar{u}$ and $\bar{p}$ are the grid-filtered velocity and pressure respectively.

The subgrid-scale eddy viscosity $\nu_{\rm SGS}$ can be computed by the Smagorinsky model
$$\nu_{\rm SGS}=(C_S\Delta)^2 (2 \overline{S}_{lk}\overline{S}_{lk})^{\frac{1}{2}}$$
where $C_S$ is the Smagorinsky coefficient and $\Delta$ is the filter width, and
$$
\overline{S}_{ij}=\overline{S}_{ij}(u)=S_{ij}(\overline{u})=\frac{1}{2}\Bigl( \frac{\partial \overline{u}_i}{\partial x_j} +  \frac{\partial \overline{u}_j}{\partial x_i} \Bigl)
$$


The LES filtered equations can then be solved e.g. by a finite element method or 
a finite volume method. 

\subsection{Vortex Model}
A vortex model is based on discretization of the vorticity field instead of the velocity field, 
\begin{equation}
    \omega=\nabla\times u. 
\end{equation}
An equation for vorticity is obtained from the Navier–Stokes equations by application of the curl operator, 
\begin{equation}
\frac{\partial \omega}{\partial t}+(u\cdot\nabla)\,u=(\omega\cdot\nabla)\,u+\nu\nabla^2\omega. 
\end{equation}

In this report free-vortex methods are used, which means that the vorticity elements are propagated with the flow velocity. The 2D model uses point vortices and the 3D model uses vortex filaments to represent the flow.
The flow velocity is obtained from the vorticity field by solving Biot-Savart's law at each time-step. For more details, see \cite{Dyachuk_2015} for more details.

\section{Validation case}
\begin{figure*}[h!]
    \centering
    \subfloat[]{
        \includegraphics[width=0.45\textwidth]{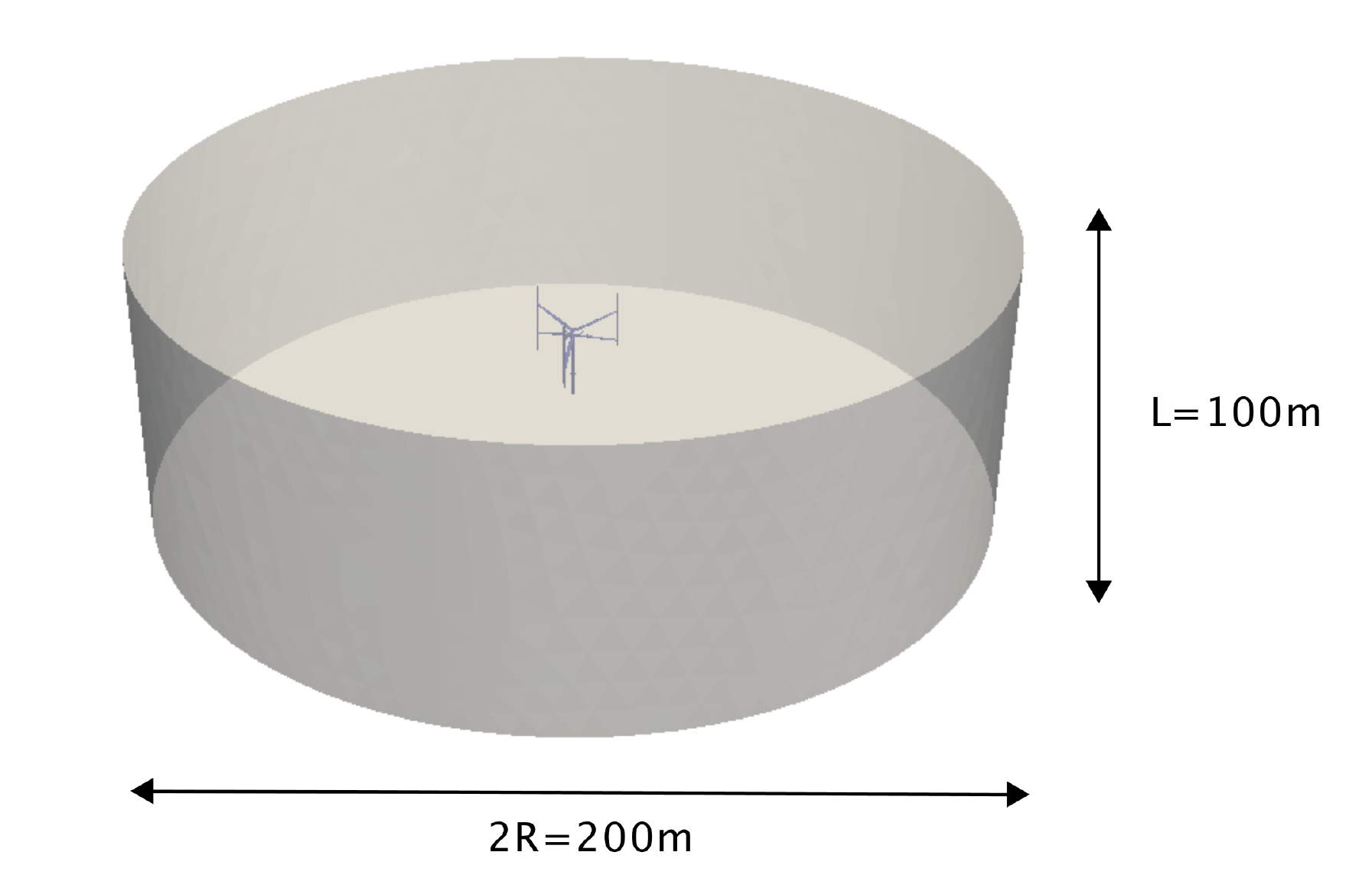}
        \label{fig:turbine_in_cylinder}
    }
    \subfloat[]{
        \includegraphics[width=0.45\textwidth]{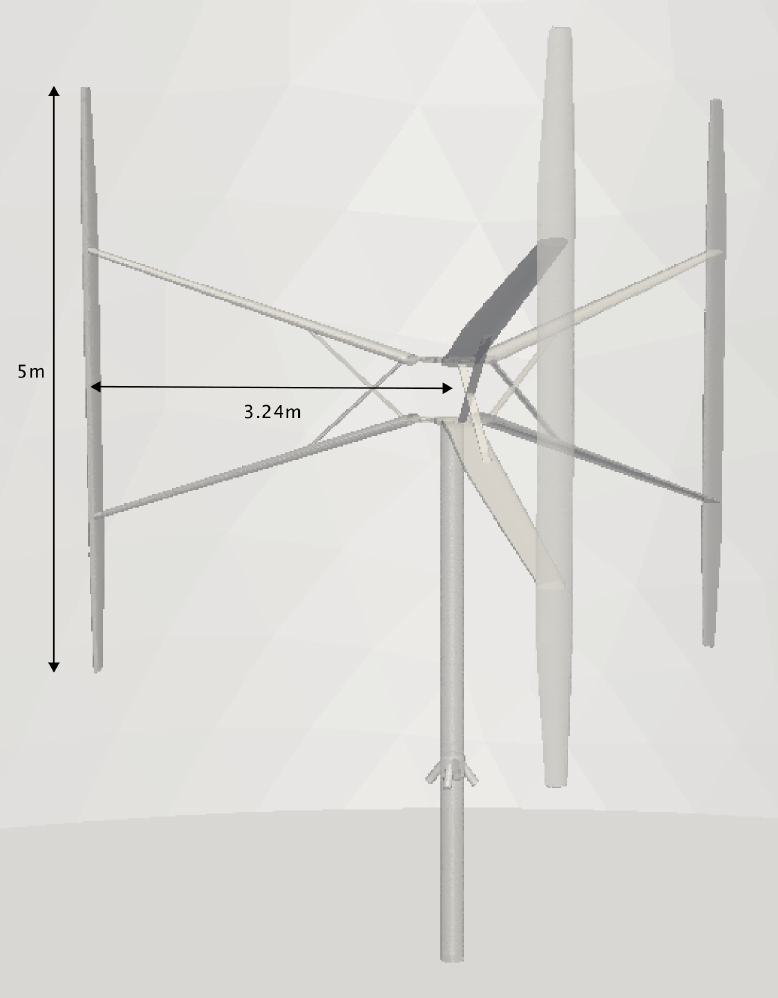}
        \label{fig:vat}
    }
    \caption{A vertical axis turbine reproduced from \cite{en81011800} placed in a cylinder. It is 1m high from the cylinder bottom to model the ground effect.}
    \label{fig:turbine3D}
\end{figure*}

The three methods (DFS, ALM and VM) were validated against experimental data in the form of blade normal forces, measured from a 12kW 3-bladed H-rotor turbine (Fig. \ref{fig:vat}) detailed in \cite{Rossander_2015,en81011800,Goude_2017, NGUYEN2019238}. The turbine radius is $r=3.24\,\rm m$ and the blade length is  ${5}\,\rm{m}$. The blades are pitched 2 degrees outwards with a chord length of ${0.25}\, \rm{m}$ at the middle of the blade. \textcolor{black}{Table \ref{tab:turbine_characteristic} gives further details of the turbine}.
\begin{table}
\centering
\caption{The turbine characteristics}
\begin{tabular}{ c c}
 \hline\hline 
 Hub height & 6 $\rm m$  \\  
 Swept area & 32 $\rm m^2$\\
 Blade airfoil & NACA0021\\
 Tapering, linear & 1 $\rm m$ (from tip)\\
 Tip chord length & 0.15 $\rm m$\\
 Mass of blade and support arms & 35.79 $\rm kg$\\
 \hline\hline   
\end{tabular}
\label{tab:turbine_characteristic}
\end{table}

For simplification, we assume that the turbine axis is coincident with
the $z-$axis and that the turbine $\Omega^\mathcal{T}$ is placed in a cylinder
$\Omega^\mathcal{C}$ with radius $R$ (Fig. \ref{fig:turbine_in_cylinder}),  
\begin{equation}
\Omega^\mathcal{C} =\Biggl\{(x,y,z)\in \bm{R}^3\Bigl| x^2+y^2\leq R^2, z\in [0; L]\Biggl\}.
\end{equation}

We set $R=100\text{m}$ and $L=100\text{m}$, which is large compared to the turbine size to minnimize artificial blockage effects. 
%
For detailed studies of the blockage effect, see \cite{GOUDE2014477,REZAEIHA2017373}.

The turbine axis coincides with the center-line of the cylinder domain, and it is placed $1 m$ above the bottom of the cylinder to model the ground effect (Fig. \ref{fig:vat}). A uniform inflow $u=(1,0,0)$ is applied on a half of the computational domain ($x<0$) and $p=0$ is applied on the other half to model an outflow boundary.

Validation studies of the ALM and VM methods was reported in \cite{Mendoza1198215}.

\section{Results}
\begin{figure*}[h!]
    \centering
    \includegraphics[width=1\textwidth]{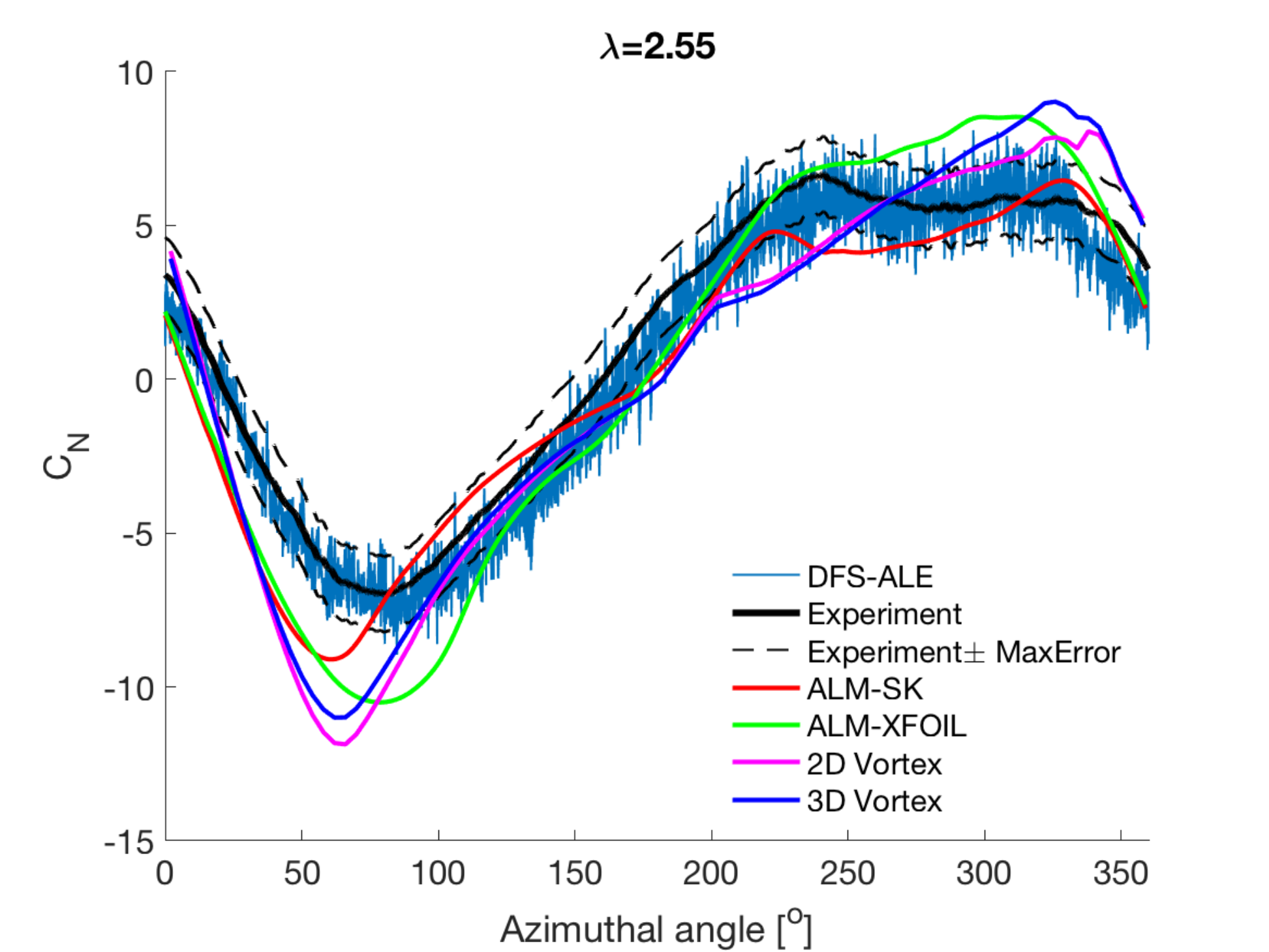}
    \caption{The normal forces for $\lambda=2.55$. The experiment was performed with $\Omega=49.89\,\rm rpm$, $U_{\infty}=6.64\,\rm m/s$.}
    \label{fig:MARSTA_lambda2p55_normalforce_cmp}
\end{figure*}

First, for a deep dynamic stall regime with tip speed ratio $\lambda=2.55$, the experiment was performed with turbine rotational speed $\Omega=49.89\,\rm rpm$ and free stream velocity $U_{\infty}=5.22\,\rm rad/s$. 
The DFS-ALE matches almost perfectly the experimental data whereas the ALM and VM methods produce lower forces between $0^\circ$ and $100^\circ$ (Fig. \ref{fig:MARSTA_lambda2p55_normalforce_cmp}). 

For the nearly optimal operational value with $\lambda=3.44$, the experiment was performed with $\Omega=64.81\,\rm rpm$ and $U_{\infty}=6.39\,\rm m/s$. 
\begin{figure*}[!h]
    \centering
    \includegraphics[width=0.9\textwidth]{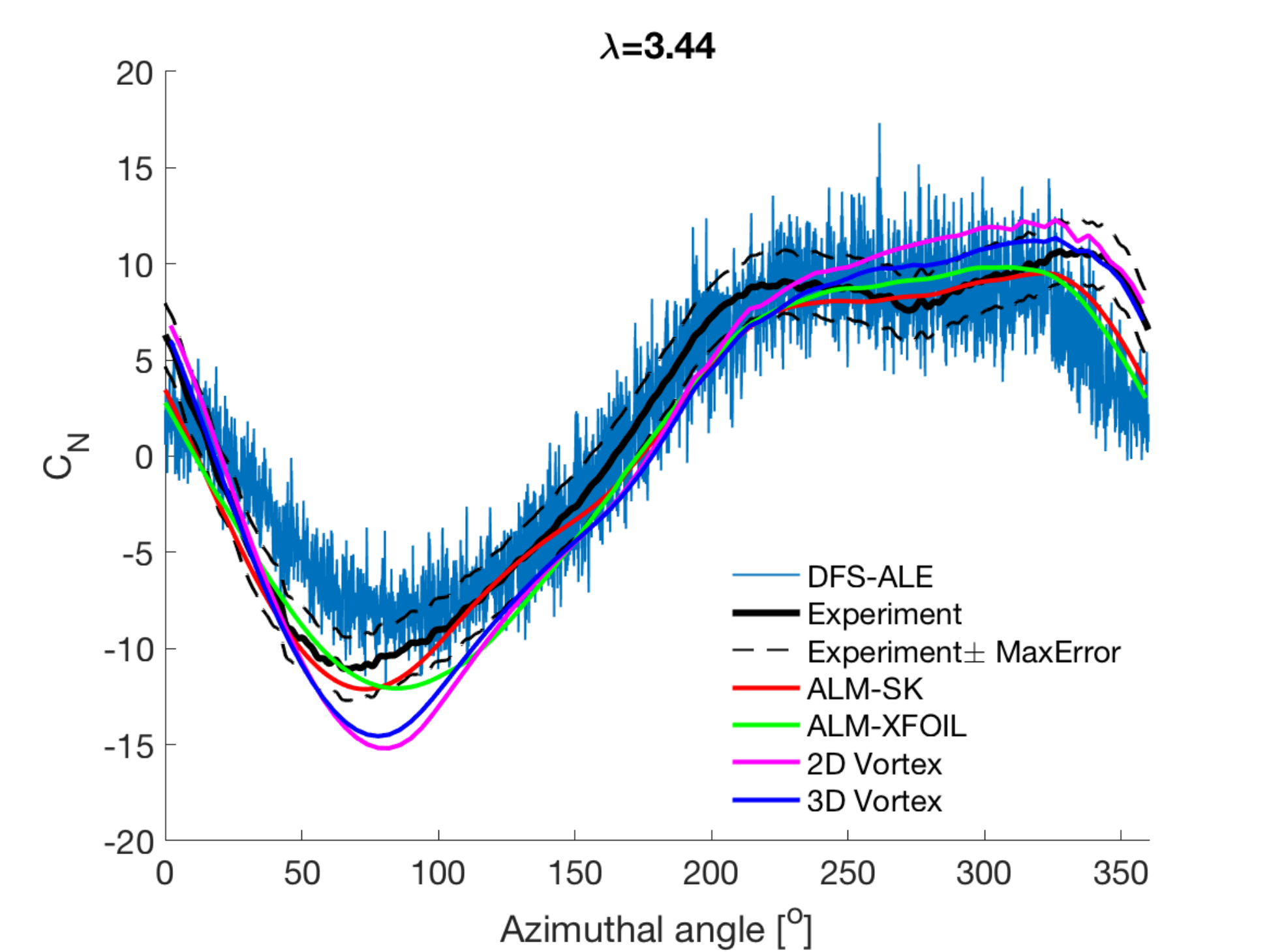}
    \caption{The normal forces for $\lambda=3.44$. The experiment was performed with $\Omega=64.81\,\rm rpm$, $U_{\infty}=6.39\,\rm m/s$.}
    \label{fig:MARSTA_lambda3p44_normalforce_cmp}
\end{figure*}

For $\lambda=4.09$, the experiment was performed with $\Omega=65.05\,\rm rpm$ and $U_{\infty}=5.39\,\rm m/s$.
\begin{figure*}[h!]
    \centering
    \includegraphics[width=0.9\textwidth]{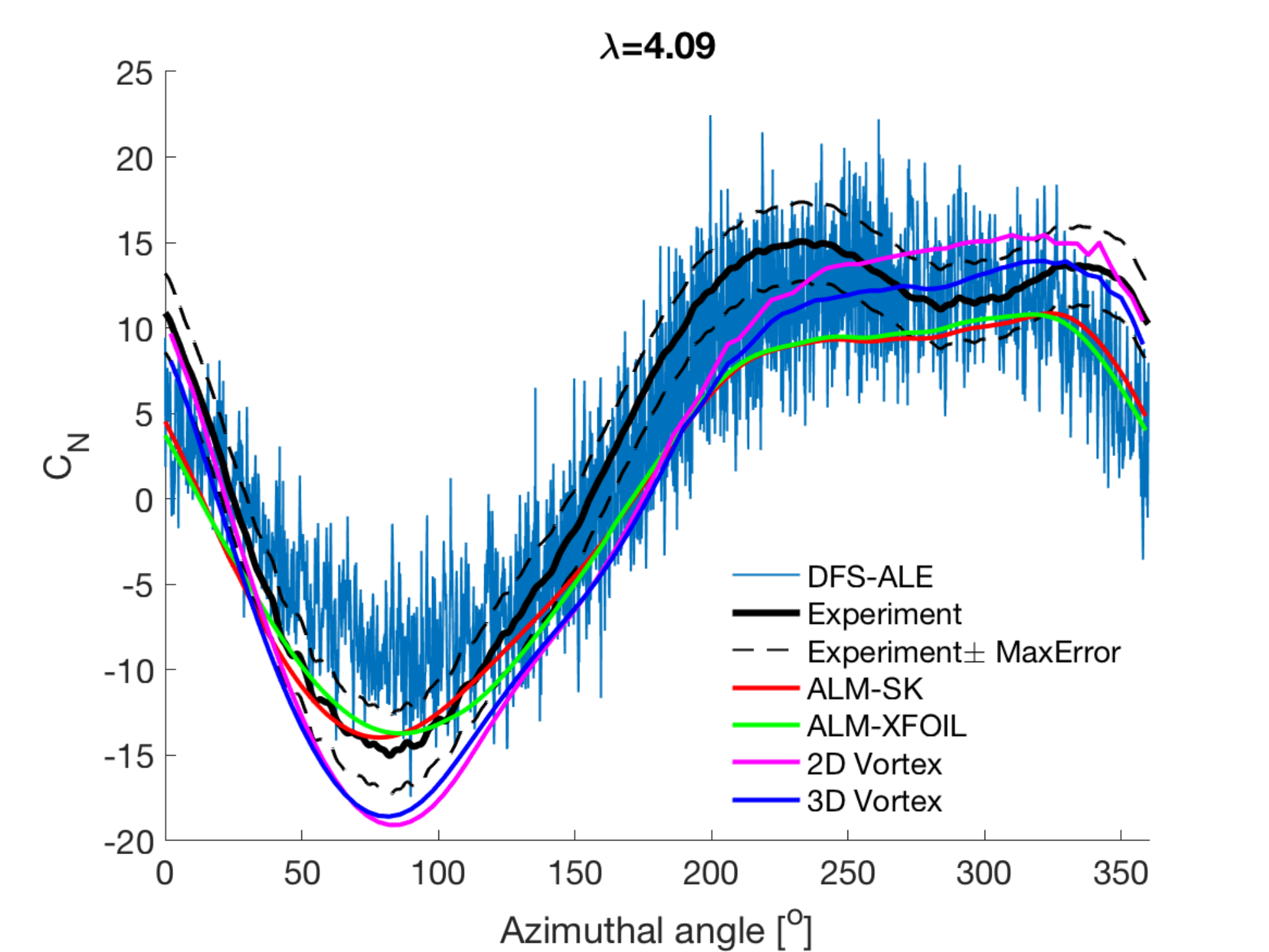}
    \caption{The normal forces for $\lambda=4.09$. The experiment was performed with $\Omega=65.05\,\rm rpm$, $U_{\infty}=5.39\,\rm m/s$.}
    \label{fig:MARSTA_lambda4p09_normalforce_cmp}
\end{figure*}
\newpage
\section*{Acknowledgement}
This research has been supported by the European Research Council, the Swedish Energy Agency, Standup for Energy, the Basque Excellence Research Center (BERC 2014-2017) program by the Basque Government, the Spanish Ministry of Economy and Competitiveness MINECO: BCAM Severo Ochoa accreditation SEV-2013-0323, the ICERMAR ELKARTEK project of the Basque Government, the Projects of the Spanish Ministry of Economy and Competitiveness with reference MTM2013-40824-P and MTM2016-76016-R.

We acknowledge the Swedish National Infrastructure for Computing (SNIC) at PDC -- Center for High-Performance Computing for awarding us access to the supercomputer resource Beskow.                                                                            
The initial volume mesh was generated with ANSA from Beta-CAE Systems S. A., who generously provided an academic license for this project.

\section*{Conflicts of Interest}
The authors declare no conflict of interest.

\section*{References}

\bibliography{myref}

\end{document}